\documentstyle[12pt,epsfig]{article}
\begin{document}
\baselineskip = 6.5mm
\topmargin= -15mm
 \begin{center}
\begin{large}
 {\bf { 
 Color triplet excitations in two dimensional QCD }}

\end{large}

\vspace{2cm}

   Makoto HIRAMOTO  and  Takehisa FUJITA

Department of Physics, Faculty of Science and Technology  
  
Nihon University, Tokyo, Japan
 
\vspace{3cm}

{\large ABSTRACT} 

\end{center}

We present a novel calculation of 
color triplet excitations in two dimensional QCD with $SU(2)$ colors. 
It is found that the lowest energy of the color triplet excitations 
is proportional to the box 
length $L$, and can be written as ${\cal M}_C={L\over{2\pi}}{g^2\over{\pi}} $.  
Therefore, the color triplet excited states go to infinity 
when the system size becomes infinity. The properties of the color triplet 
states such as the wave functions are studied for the finite box length.

\vspace{1cm}
\noindent
PACS numbers: 11.10.Kk, 03.70.+k, 11.30.-j, 11.30.Rd

\newpage

\begin{enumerate}
\item{\large Introduction}

In Quantum Chromodynamics (QCD), physical particles are all color singlet 
states. The quarks cannot be free, and there is no color excitation. 
This is a well known fact, and there is no doubt about it. 

The confinement mechanism in QCD plays a most important role, and 
one learns in textbooks that the quark and antiquark cannot be 
separated since they feel linear confining forces between them 
when they are apart. Therefore, there are no free quarks in nature, 
and this must be of course a right picture for QCD. 

However, this may not be everything for the confinement mechanism 
of QCD. For example, bosonic states with color octet in $SU(3)$ color 
should not exist in nature. But naive questions are in which way 
they cannot exist and why they should not be found in nature. 
At least, a simple-minded picture suggests that they have little to do 
with the linear confining potential since the non-existence of the bosonic 
states with colors should not necessarily be due to the confining force 
between the quarks.  Instead, the bosonic objects themselves 
with colors cannot exist. 

In order to understand physics in QCD in depth, we study the behavior 
of the color triplet excitations in QCD$_2$ 
with $SU(2)$ color. The calculation of the color triplet 
excitations in QCD$_2$ is carried out with the Bogoliubov vacuum state \cite{q11,q10,q9}, 
and we find that the lowest state energy of the color triplet excitations ${\cal M}_C$ 
can be described as 
$$ {\cal M}_C={L\over{2\pi}}{g^2\over{\pi}}  \eqno{(1.1)} $$
where $L$ denotes the box length, and $g$ is the gauge coupling constant. 
Therefore, these color triplet excitations are not realized in nature 
since we should let the box length $L$ infinity at the final stage. 

This is in contrast to the singlet boson in QCD$_2$ 
with $SU(N)$ colors \cite{q01,q211,q21,q210}. 
In a recent paper \cite{qq1}, the mass of the boson 
 ${\cal M}_{N}$ is expressed for large values of $N$ 
at the massless fermion limit, 
$$  {\cal M}_{N}={2\over 3}\sqrt{{Ng^2\over{3\pi}}} . \eqno{(1.2)} $$
This boson mass together with the singlet excitations do not depend 
on the box length $L$ at all, once the box length $L$ is taken to be 
sufficiently large. Therefore, it is clear that these states 
can be realized in nature.

However, as long as the box length is finite, the color triplet 
excitations have of course some finite excitation energy, 
and from the study of the behavior of the wave function, 
we may learn the basic mechanism of the non-existence of 
the color triplet excitations in QCD$_2$. 

In this paper, we clarify the behavior of the wave function of the color triplet state, 
and show that the color triplet state has a strongly localized component 
in momentum space, and therefore, in coordinate space, it should 
be rather flat. Therefore, the color triplet state has a very small 
kinetic energy, but the potential energy from interactions 
gives rise to the excitation energy which is proportional to the box length $L$. 

Here, we present a schematic explanation of the energy behavior 
of the color singlet boson and color triplet excitation. 
The total energy of the bosonic states for QCD$_2$ can be schematically written
$$ {\cal E}= {a\over L}+bLg^2 \eqno{(1.3)} $$
where $a$ and $b$ are simple dimensionless constants. 
The first term corresponds to the 
kinetic energy term, and the second term is due to the interactions.  

For singlet bosonic states, we minimize 
the energy $\cal E$, and obtain the boson mass
$$ {\cal M}= 2\sqrt{ab} g .  \eqno{(1.4)} $$
Obviously, the singlet bosonic states do not depend on the box length $L$. 
But it should be noted that the minimum energy of eq.(1.4) can be realized 
only when the two terms are just in the same magnitude. 

On the other hand, the color triplet states have very large repulsive 
energy from the interaction part, and therefore, the two terms 
cannot become the same order of magnitude. 
From the calculations, we know that the color triplet states 
have strong localizations around $p=0$ in momentum space, 
and therefore the color triplet states 
behave like the energy given in eq.(1.1).

\vspace{1cm}

\item{\large Mass of color singlet boson in QCD$_2$}

In this section, we summarize the calculated results 
of the boson mass and the condensate value of 
the color singlet states in QCD$_2$. 
The detailed discussions are found in \cite{qq1}. 
The calculations are carried out with the Bogoliubov vacuum, 
and it is shown that the calculated values of the condensate 
agree very well with prediction of the $1/N$ expansion. 
Also, the calculated boson mass is finite at the massless fermion 
limit, and this is quite consistent with the constraint 
that is inherent to the two dimensional field theory. Namely, 
the massless boson should not physically exist in nature. 

In Table 1, we summarize the results obtained in \cite{qq1}. 

\begin{center}
Table 1 \\
\ \ Boson mass ${\cal M}_{N}$  and condensate $C_{N}$ in  QCD$_2$ \\ 
\ \ \\
\begin{tabular}{|c||c|c|}
\hline
\  & \  & \  \\
$SU(N)$ & ${\cal M}_{N}$ & $C_{N}$   \\
\  & \  & \  \\
\hline
\hline
\  & \  & \  \\
$N=2$ & \  0.467${g\over{\sqrt{\pi}}}$ & 
$\  -0.495{g\over{\sqrt{\pi}}}$  \\
\  & \  & \  \\
\hline
\  & \  & \  \\
$N=3$ & \  0.625${g\over{\sqrt{\pi}}}$ & 
\  $-0.995{g\over{\sqrt{\pi}}}$  \\
\  & \  & \  \\
\hline
\  & \  & \  \\
$N>>1$ &\   $ {2\over 3}\sqrt{{Ng^2\over{3\pi}}}$ & 
\  $ -{N\over{\sqrt{12}}} \sqrt{Ng^2\over{2\pi}} $  \\
\  & \  & \  \\
\hline
\end{tabular} 
\end{center}

\vspace{1cm}

\item{\large   Bogoliubov transformation in QCD$_2$ }

In this section, we briefly discuss the Bogoliubov transformation in QCD$_2$. 
The Lagrangian density for QCD$_2$ with $SU(N)$ color is described as 
$$ {\cal L}= 
\bar{\psi}(i\gamma^{\mu}\partial_{\mu}-g\gamma^{\mu}A_{\mu}-m_0 )\psi 
-\frac{1}{4}G^{a}_{\mu\nu}G^{a\mu\nu},
 \eqno{(3.1)} $$
where $G_{\mu\nu}$ is written as
$$ G_{\mu\nu} = \partial_{\mu}A_{\nu}-\partial_{\nu}A_{\mu}+ig[A_{\mu},A_{\nu}] $$
$$
A_{\mu} = A^{a}_{\mu}T^{a}, \ \ \  
T^{a} = \frac{\tau^{a}}{2} .  $$

Now, we first fix the gauge by 
$$ A^a_1 =0 . \eqno{(3.2)} $$
In this case, the Hamiltonian of QCD$_2$ with $SU(N)$ color  can be written as 
$$ H = \sum_{n,\alpha}p_{n}
\left(a_{n,\alpha}^{\dagger}a_{n,\alpha}-b_{n,\alpha}^{\dagger}b_{n,\alpha}
\right)  +m_0 \sum_{n,\alpha}
\left(a_{n,\alpha}^{\dagger}b_{n,\alpha}+b_{n,\alpha}^{\dagger}a_{n,\alpha}
\right) $$
$$ -\frac{g^2}{4NL}\sum_{n,\alpha,\beta}\frac{1}{p^{2}_{n}}
\left({\tilde j}_{1,n,\alpha\alpha}+{\tilde j}_{2,n,\alpha\alpha}
\right)
\left({\tilde j}_{1,-n,\beta\beta}+{\tilde j}_{2,-n,\beta\beta}
\right)
\nonumber $$
$$ +\frac{g^2}{4L}\sum_{n,\alpha,\beta}\frac{1}{p^{2}_{n}}
\left({\tilde j}_{1,n,\alpha\beta}+{\tilde j}_{2,n,\alpha\beta}
\right)
\left({\tilde j}_{1,-n,\beta\alpha}+{\tilde j}_{2,-n,\beta\alpha}
\right) \eqno{(3.3)} $$
where
$$ 
{\tilde j}_{1,n,\alpha\beta} = 
\sum_{m}a_{m,\alpha}^{\dagger}a_{m+n,\beta} \eqno{(3.4a)} $$
$$ {\tilde j}_{2,n,\alpha\beta} = 
\sum_{m}b_{m,\alpha}^{\dagger}b_{m+n,\beta} . \eqno{(3.4b)} $$
Now, we define new fermion operators by the Bogoliubov transformation, 
$$ a_{n,\alpha}=\cos\theta_{n,\alpha}c_{n,\alpha}+\sin\theta_{n,\alpha}
d_{-n,\alpha}^{\dagger} \eqno{(3.5a)} $$
$$
b_{n,\alpha}=-\sin\theta_{n,\alpha}c_{n,\alpha}+\cos\theta_{n,\alpha}
d_{-n,\alpha}^{\dagger} \eqno{(3.5b)} $$
where $ \theta_{n,\alpha}$ denotes the Bogoliubov angle. 

In this case, the Hamiltonian of QCD$_2$ can be written as 
$$ H = \sum_{n,\alpha}E_{n,\alpha}(c^\dagger_{n,\alpha}c_{n,\alpha}
+d^\dagger_{-n,\alpha}d_{-n,\alpha}) +H' \eqno{(3.6)} $$
where
$$ E^2_{n,\alpha} = \left\{p_{n}+\frac{g^2}{4NL}\sum_{m,\beta}
\frac{(N\cos 2\theta_{m,\beta}-\cos 2\theta_{m,\alpha})}{(p_m-p_n)^2}
\right\}^2 $$
$$ +\left\{m_0+ \frac{g^2}{4NL}\sum_{m,\beta}{
(N\sin 2\theta_{m,\beta}-\sin 2\theta_{m,\alpha}) 
\over{{(p_m-p_n)^2} }} \right\}^2 .
 \eqno{(3.7)} $$
$H'$ denotes the interaction Hamiltonian in terms of the new operators 
and is given in Appendix of \cite{qq1}. 

Now, we can calculate the boson mass for the $SU(3)$ color. Here,  we 
define the wave function for the color triplet boson as 
$$ |\Psi_{0}\rangle = \frac{1}{\sqrt{2}}\sum_{n}f_{n}
(c^\dagger_{n,1}d^\dagger_{-n,1}-c^\dagger_{n,2}d^\dagger_{-n,2})
|0\rangle .  \eqno{(3.8)} $$
In this case, the boson mass $ {\cal M}_C$ can be described as 
$$ {\cal M}_C = \langle \Psi_{0}|H|\Psi_{0}\rangle
 = \sum_{n,\alpha=1,2}E_{n,\alpha}|f_n|^2 $$
$$
-\frac{g^2}{8L}\sum_{l,m}\frac{f_{l}f_{m}}{(p_l-p_m)^2}
\Big(\cos(\theta_{l,1}-\theta_{m,1})
\cos(\theta_{l-K,1}-\theta_{m-K,1})$$
$$+\cos(\theta_{l,2}-\theta_{m,2})
\cos(\theta_{l-K,2}-\theta_{m-K,2})\Big) $$
$$+\frac{g^2}{4L}\sum_{l,m}\frac{f_{l}f_{m}}{(p_l-p_m)^2}
\Big(\cos(\theta_{l,1}-\theta_{m,2})\cos(\theta_{l-K,1}-\theta_{m-K,2})$$
$$+\cos(\theta_{l,2}-\theta_{m,1})\cos(\theta_{l-K,2}-\theta_{m-K,1})\Big) $$
$$ -\frac{g^2}{8L}
\sum_{l,m}\frac{f_{l}f_{m}}{K^2}
\Big(\sin(\theta_{l,1}-\theta_{l-K,1})
\sin(\theta_{m-K,1}-\theta_{m,1})$$
$$+\sin(\theta_{l,2}-\theta_{l-K,2})
\sin(\theta_{m-K,2}-\theta_{m,2})$$
$$+\sin(\theta_{l,1}-\theta_{l-K,1})
\sin(\theta_{m-K,2}-\theta_{m,2})$$
$$+\sin(\theta_{l,2}-\theta_{l-K,2})
\sin(\theta_{m-K,1}-\theta_{m,1})\Big)  \eqno{(3.9)} $$
where the last terms are finite with $K$ $\rightarrow 0$. 
This equation can be easily diagonalized together with the Bogoliubov 
angles, and we obtain the boson mass as the function of $L$.

\vspace{1cm}

\item{\large   Color triplet excitation}

We carry out the calculations of eq.(3.9), and obtain 
the  spectrum of the color triplet 
excited states with the massless fermion. In Fig. 1, we show the first excited state 
of the color triplet states as the function of the box length $L$ 
and the matrix dimension $D$. 
The solid line is the phenomenological formula of eq.(1.1),
$$ {\cal M}_C={L\over{2\pi}}{g^2\over{\pi}} .  \eqno{(1.1)} $$
This fits to the calculated spectrum quite well. 

Eq.(1.1) indicates that the color triplet excitation becomes infinity 
when the box length $L$ becomes infinity, which is the real nature. 
Therefore, this state cannot be observed in nature. On the other hand, 
the singlet boson mass for $SU(2)$ is determined to be \cite{qq1}
$$ {\cal M}=0.467{g\over{\sqrt{\pi}}}  \eqno{(4.1)} $$
which does not depend on the box length $L$ at all, and therefore  
this boson can certainly be observed in nature. 

\begin{center}
\begin{figure}
\includegraphics*[width=9cm, height=6cm]{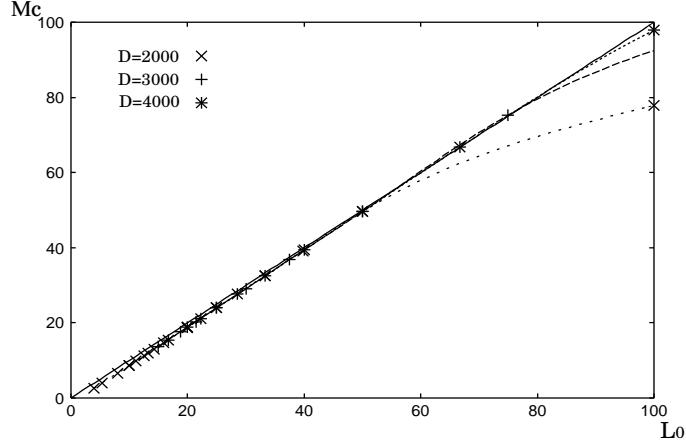}
\caption{The lowest excitation energy  of the color triplet bosonic 
state is shown as the function of the box length $L_0={L\over{2\pi}}$. }
\end{figure}
\end{center}

\begin{figure}
\includegraphics*[width=9cm, height=6cm]{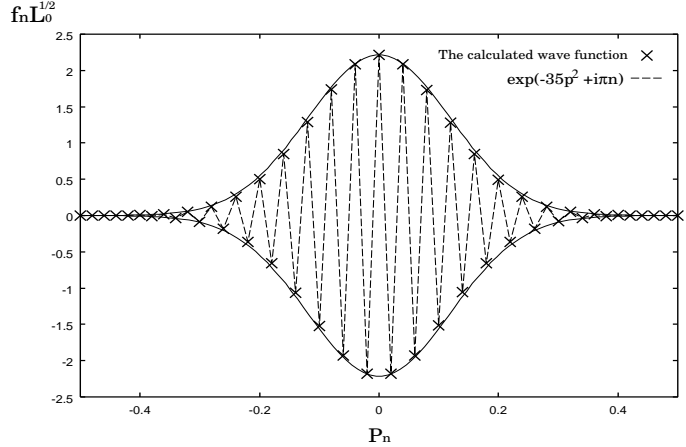}
\caption{The wave function in momentum space of the color triplet bosonic 
state is shown as the function of $p_n={2\pi n\over{ L}}$. 
Here, we take $D=4000$. }
\end{figure}

However, as long as the box length $L$ is finite, the color triplet 
excitation energy is finite and the wave function can be determined. 
In Fig. 2, we show the calculated wave function $f_n$ for the first 
excited state of the color triplet states. Here, $L_0={L\over{2\pi}}$ is taken to be 
$L_0=50/{g\over{\sqrt{\pi}}}$, and the matrix dimension is $D=4000$. 
The wave function for this case can be phenomenologically described as 
$$ f_n = f_0 \exp \left(-  35\left({p_n\over{{g\over{\sqrt{\pi}}}}}\right)^2
+i\pi n \right) \eqno{(4.2)} $$
Therefore, the wave function plotted in Fig. 2 is found only 
around $p_n=0$ area, and it is quite localized in momentum space.  
Therefore, it should be flat in coordinate 
space.

In Fig. 3, the energy spectrum of the singlet as well as the triplet states 
in QCD$_2$ are shown. Here, the box length $L$ is fixed 
to $L_0=50 /{g\over{\sqrt{\pi}}}$.  
As can be seen, the triplet excitation energies are much higher 
than those of the singlet states. 
As stated above, these color  singlet bosonic states  do not depend on the box 
length $L$ at all. 
In this respect, the color triplet states are very special. 

\begin{figure}
\includegraphics*[width=9cm, height=6cm]{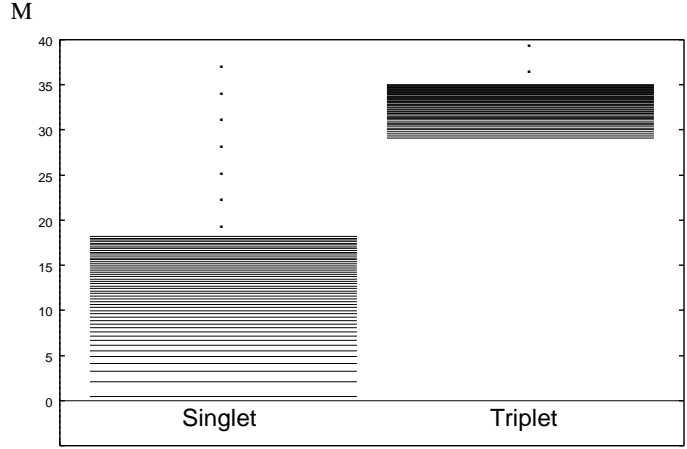}
\caption{We show the calculated spectrum of the color  singlet and triplet 
bosonic states}
\end{figure}

\vspace{1cm}

\item{\large   Intuitive interpretation of color triplet excitations}

What is the physics behind the difference between the triplet 
excitations and the singlet bosonic states ? 

In order to intuitively understand the physics in connection 
to the confinement mechanism, we carry out the evaluation 
of the interaction term which becomes dominant to the color 
triplet state. This dominant  part for the color triplet states 
is the last term of eq.(3.9). It is interesting to note that 
the last term of  eq.(3.9) completely vanishes for the singlet state 
due to the kinematical constraint of $SU(2)$ group. 
On the other hand, the color triplet states are quite 
different, and the last term of eq.(3.9) gives rise to 
the very large repulsive  energy for the color triplet states. 

Further, we can easily show that the last term of eq.(3.9) 
cannot be described as the local potential, 
but it is a highly nonlocal 
interaction. In fact, the last part of eq.(3.9) can be approximately 
described in terms of 
the separable interaction in momentum space,
$$ H_{nm}= {V_0 \over{ \left( ({n\over a_0})^2 +1\right) 
\left( ({m\over a_0})^2 +1\right) }  } f_nf_m \eqno{(5.1)} $$
where $V_0$ and $a_0$  are some constants which depend on the 
values of $D$ and $L$. 

Due to the nonlocality of the interaction, 
it gives rise to the energy which is proportional to the box 
length $L$. Even for the nonrelativistic kinematics, 
one cannot find a behavior which is expected from a simple-minded potential 
picture for the interaction term of the last term in eq.(3.9).

\vspace{1cm}

\item{\large   Conclusions}

We have presented a novel calculation of the bosonic excitation energy 
for the color triplet state in QCD$_2$ with $SU(2)$. 
The bosonic excitation energy is proportional 
to the box length $L$, and therefore it goes to infinity when 
the box length $L$ becomes infinity. This means that the color triplet 
excited states cannot exist in nature. This nonexistence of the color 
excited states seems to have little to do with the confinement potential, 
but it is the consequence of the $SU(2)$ kinematics rather than the potential 
shape like the linear rising potential. Further studies on this line 
may well be a good help to clarify the color confinement mechanism 
in four dimensional QCD.

\end{enumerate}
\end{document}